# Approximation of the reception coefficients of cosmic rays neutron component for latitude measurement

**Kobelev P.G., Yanke V.G.**

*The Pushkov Institute of Terrestrial Magnetism, Ionosphere and Radiowave Propagation of the Russian Academy of Sciences (IZMIRAN), Moscow, RF*     yanke@izmiran.ru, kobelev@izmiran.ru

**Abstract.** When solving scientific and applied problems, such as latitude monitoring, it is important to correctly exclude primary cosmic ray variations from observation data. Therefore, the purpose of this study was to develop and implement a method for correcting monitoring data, the key point of which was to obtain reception coefficients as a function of latitude. This resulted in an approximation of the rigidity dependence of the zeroth and first harmonic cosmic rays anisotropy coefficients, calculated for a ground-based cosmic ray detectors network. Analysis of the obtained results showed that the approximation was performed with high accuracy, and the results are suitable for use in latitude measurements during marine expeditions.

## 1. Introduction

Galactic cosmic rays (CR) reaching Earth almost isotropically. However, a small (several percent) anisotropy of the CR intensity angular distribution, caused by dynamic processes in the solar wind, constantly exists and exceeds stellar anisotropy by orders of magnitude. The bulk of this anisotropy is due to the Earth's rotation around its axis. This is the solar-diurnal anisotropy of the first harmonic, which manifests itself in observational data as a diurnal wave in CR intensity. This component can be represented as a vector and is called vector anisotropy, reflecting the CR current. The current is determined by two vector components: convective and diffusive. The convective current arises due to the features of the CR motion in the interplanetary magnetic field and is directed away from the Sun. The diffusive current is caused by the CR gradient and their scattering by magnetic inhomogeneities and is directed toward the Sun along the interplanetary magnetic field lines. But often, especially during periods of strong modulation, a more complex angular distribution of CR is also observed – the second harmonic of anisotropy [Gololobov et al., 2017].

Over nearly a century of CR anisotropy research, a large number of papers have been published (a list can be found in [Belov et al., 2023]). Constant attention to this problem is due to the fact that anisotropy is sensitive to all major manifestations of solar activity and, accordingly, is also sensitive to all modulation processes in the heliosphere.

Local and global analysis methods have been developed to get CR anisotropy [Belov et al., 2023]. The popularity of CR anisotropy studies using local methods has been facilitated by the simplicity of getting it from measurement data. However, local methods for getting anisotropy (e.g., the superimposed epoch method or the harmonic analysis method) require stationary data series and have several disadvantages: lower statistical accuracy and possibility of usage only during relatively quiet periods of solar activity (SA). Furthermore, local methods require a separate procedure for analyzing variations in the heliosphere beyond the Earth's atmosphere and magnetosphere.

Global Survey Method for primary CR variations studying combines simultaneous ground-based observations of CR at all detectors of the network and, according to G.F. Krymsky, is a more sophisticated version of spherical analysis that takes into account the Earth's atmosphere and magnetosphere. The method allows us to determine the CR density (the isotropic part of the intensity) and CR anisotropy for each hour outside the magnetosphere and atmosphere and makes it possible to study rapid anisotropy changes associated with the dynamics of the interplanetary medium. It essentially combines three methods: the coupling coefficient method [Dorman, 1957], which takes into account the atmosphere; the method of trajectory calculations of particle motion in the geomagnetic field [McCracken et al., 1962; Dorman et al., 1971], which takes into account the Earth's magnetosphere; and the method of spherical analysis [Korn & Korn, 1974], which makes it possible to factor out spherical harmonics for further analysis.

The multi-channel nature of such a device provides reliable and continuous measurements, as the result the statistical accuracy of the network increased significantly and the influence of instrumental effects significantly reduced. For example, a single neutron monitor (18NM64) at sea level provides a statistical accuracy of ~0.15%/h, while the entire network of stations provides an accuracy of approximately ~0.01%/h.

One of the earliest and most successful implementations of the global method was developed in Yakutsk [Krymskiy et al., 1966a, 1966b, 1967, 1981]. Around the same time, Japanese researchers proposed their own methodology [Nagashima, 1971]. Various modifications of this method were later developed, for example, in [Belov et al., 1973, 1974, 2018].

A group of Irkutsk researchers developed their own version of the global spectrographic method [Dvornikov & Sdobnov, 1998], which has become the most complex and multipurpose of the existing ones. It is used to calculate the first and second harmonics of CR anisotropy, to determine variations in the planetary distribution of geomagnetic cutoff rigidities, and to estimate the variations spectrum.

When solving applied tasks, it is often necessary to exclude cosmic ray variations expected at a given location from detector observation data, which is an actual problem. In the isotropic approximation, this can be accomplished using reference station data [Kobelev et al., 2022]. In first harmonic approximation, the expected variations can be estimated based on GSM results. IZMIRAN has implemented the GSM method for the period from 1957 to the present, and the results of this analysis can be found on Yandex Disk [https://disk.yandex.ru/d/mKHMM2dztqNoHw (GSM section)] or on the website [https://crst.izmiran.ru/w/feid]. Therefore, using the GSM results for estimating variations in the first harmonic approximation, it is necessary to solve the direct problem (finding the expected variations at a given point and time). However, to solve this problem, it is necessary to know the reception coefficients, which were calculated in [Yasue et al., 1982] only at discrete points for the detector network. Therefore, it is necessary to perform laborious calculations at arbitrary points along the vessel's route or to approximate the reception coefficients.

The objective of this work is to approximate the reception coefficients of the first harmonic, which is carried out for the first time. This will allow to evaluate the reception coefficients for newly established neutron monitoring stations, as approximately 20 new neutron detectors have been built since the publication of [Yasue et al., 1982]. Furthermore, this will allow for processing latitudinal measurement data from marine expeditions.

The aim of this work is to develop and implement a method for correcting latitude measurement data for primary CR variations. To achieve this, the primary objective is to approximate the zero and first harmonic reception coefficients as a function of geomagnetic cutoff rigidity.

## 2. Global survey method: direct and inverse problems

Let's briefly outline the principle of the GSM method and introduce the necessary notations. Let's assume that the solution of the inverse problem, namely, the zero-order amplitude and the vector of the first spherical harmonic $\xi^{GEO} = (\xi_x^{GEO}, \xi_y^{GEO}, \xi_z^{GEO})$ is known.

The characteristics are defined in the geophysical system GEO, in which the X-axis is directed from the Earth to the Sun, and the Y-axis is located at the equatorial plane and is directed towards the sun (opposite to the rotation of the planet).

We present an algorithm for solving the direct problem, i.e., the problem of determining variations at given points. The coordinates of ground-based detectors are also defined in the geographic coordinate system (**GEO**), the $X'$ axis of which passes through the Greenwich prime meridian in the Earth's equator plane. The $Z'$ axis is parallel to the Earth's rotation axis. To determine the vector of the first spherical harmonic $\xi' = (\xi_x', \xi_y', \xi_z')$ in the terrestrial coordinate system associated with the prime meridian, it is necessary for each i-th detector ($i=1..m$) to perform a rotation around an axis by an angle $\varphi_t^i = 2\pi/24(t+½)$, which corresponds to $t$ hours

$$\xi' = M_Z^i \cdot (\xi^{GEO})^T, \tag{1}$$

where $M_Z^i = \begin{pmatrix} \cos\varphi_t^i & \sin\varphi_t^i & 0 \\ -\sin\varphi_t^i & \cos\varphi_t^i & 0 \\ 0 & 0 & 1 \end{pmatrix}. \tag{2}$

Then the expected CR intensity variations on Earth for each observation point, in the designations by [Yasue et al., 1982], can be written as

$$v^i = aC_{00}^i + C_x^i \xi_x^{GEO} + C_y^i \xi_y^{GEO} + C_z^i \xi_z^{GEO}, \tag{3}$$

where $C_{00}^i$ and $C^i = (C_x^i, C_y^i, C_z)$ are the reception coefficients linking variations outside the magnetosphere with variations observed on the Earth. These coefficients depend on the coordinates and observation altitude for each detector located at a given point. The reception coefficients calculated in [Yasue et al., 1982] are given in a polar coordinate system, and to convert to a Cartesian coordinate system, it is necessary to carry out the following transformations for each station, taking into account longitude $\varphi_s$ and drift angle $\varphi_{11}$ of particles in the Earth's magnetic field

$$\begin{aligned} C_x^i &= -A_{11}^i Cos(\varphi_S^i + \varphi_{11}^i) \\ C_y^i &= -A_{11}^i Sin(\varphi_S^i + \varphi_{11}^i) \\ C_z^i &= -C_{10}^i \end{aligned} \tag{4}$$

where $A_{11}^i$ is the amplitude of the first spherical harmonic of the anisotropy equatorial component and $\varphi_{11}^i$ – the effective angle of particle drift in the Earth's magnetosphere according to data [Yasue et al., 1982]. The reception coefficients of the north-south asymmetry and the reception coefficient of zero harmonic $C_{00}^i = \int_{R_c^i}^{R_U} W^i(R) \cdot R^{-\gamma} dR$ are also calculated. The choice of sign in (4), and accordingly the coordinate system, is determined by the characteristic with which it is intended to work – the anisotropy vector or the cosmic ray current.

For the direct problem in matrix form, expression (3) in the case of the first harmonics can be written as

$$v^i = aC_{00}^i + C^i M_Z^i (\xi^{GEO})^T, \tag{5}$$

or, expanding (5) with taking into account (2) and (4), for the direct problem we obtain

$$v^i(t) = aC_{00}^i + C_x^i(\xi_x^{GEO}\cos\varphi_t^i + \xi_y^{GEO}\sin\varphi_t^i) + C_y^i(-\xi_x^{GEO}\sin\varphi_t^i + \xi_y^{GEO}\cos\varphi_t^i) + \xi_z^{GEO}C_z^i, \tag{6}$$

i.e., according to the known zero harmonic amplitude *a* and the vector of the first spherical harmonic $\xi^{GEO} = (\xi_x^{GEO}, \xi_y^{GEO}, \xi_z^{GEO})$, CR variations at the point *i* with longitude $\varphi_S^i$ at time *t*, which is defined as $\varphi_t^i = \varphi_S^i + \varphi_{11}^i$, are estimated.

For the inverse problem, it is necessary to transform the system of equations (6) as

$$v^i(t) = aC_{00}^i + \xi_x^{GEO}(C_x^i \cos\varphi_t^i - C_y^i \sin\varphi_t^i) + \xi_y^{GEO}(C_y^i \cos\varphi_t^i + C_x^i \sin\varphi_t^i) + \xi_z^{GEO}C_z^i, \qquad (7)$$

and as a result we get a linear system with *m* equations relatively variables *a* and $\xi^{GEO}$ in the geocentric solar-ecliptic **GEO** coordinate system.

The results of such an analysis (7) for hourly resolution for the period from 1957 to the present can be found on the Yandex Disk resource https://disk.yandex.ru/d/mKHMM2dztqNoHw (GSM part) or https://crst.izmiran.ru/w/feid.

## 3. Used data

The reception coefficients for discretely located neutron monitors of the world network were calculated in [Yasue et al., 1982], and for muon telescopes – in [Fujimoto et al., 1984]. These data can be found on the Yandex Disk resource [https://disk.yandex.ru/d/mKHMM2dztqNoHw (GSM part)]. The reception coefficients of zero, first and second harmonics were calculated for several values of the power indices γ of variations spectrum and several values of upper rigidity $R_U$ of CR modulation in geliosphere.

It should be noted that reception coefficients were calculated for the magnetosphere of the 1975 epoch. Over the past 50 years the magnetosphere was restructured [Gvozdevsky et al., 2016], which should leads to changing of asymptotic directions and of geomagnetic cutoff rigidities. Two anomalous zones exist: the South Atlantic, with an accelerated decrease of $R_c$ (to -2 *GV*), and the North Atlantic, with an increasing $R_c$ (to +2 *GV*). During the period of the cosmic ray observations, the planetary-average geomagnetic cutoff rigidity for vertical particle arrivals decreased by 0.2 *GV*.

## 4. Sensitivity of neutron monitors to isotropic and anisotropic variations

An important characteristic of a neutron monitor is its sensitivity to various types of variations. When constructing a network of detectors, it is important to determine which ones are the most sensitive — that is, which ones respond most effectively to various types of variations and are best suited for solving specific problems.

To quantitatively the sensitivity to isotropic variations $C_{00}$ by using the CR coupling function $W(R, h_0)$ and spectrum of primary variations (in simplest case, purely power-law $R^{-\gamma}$) as

$$C_{00}(R_c) = \int_{R_c}^{R_U} W(R, h_0)(R/10GV)^{-\gamma} dR, \qquad (8)$$

where $R_U$=100 ГВ - upper rigidity of the modulation area.

**Figure** 1 (left panel) shows the zero harmonic reception coefficients of the world network of neutron monitors for several indices of the power-law variations spectrum according to the data of [Yasue et al., 1982]. For comparison, the muon component reception coefficients of two vertical muon telescopes, Nagoya and Yakutsk, are also shown [Fujimoto et al., 1984]. As expected, the most sensitive are the high-altitude polar detectors. Indeed, for such detectors, at γ=0.5 (characteristic value for Forbush - decrease) reception coefficient $C_{00}$ is ~1 %/%, i.e. isotropic variation 1 % outside the magnetosphere will lead to corresponding observed variations on Earth. The reception coefficients $C_{00}$ are also given for γ=1.5 (characteristic value for long-term variations at SA minimums with *A*<0) and for γ=1 (characteristic value for long-term variations at SA minimums with *A*>0 and at all SA maximums [Yanke et al., 2019]).

For a soft rigidity variations spectrum (an example for γ =1.5, 1.0) the zero harmonic reception coefficient for some mountain detectors may be greater than 1. The estimate of the reception coefficients is given for the variation spectrum normalized to an effective rigidity 10 GV. In reality, depending on the soft spectrum of variations and maximum rigidity in the modulation area $R_U$, such effective particle rigidity is not achieved, for example, for all mountain stations.

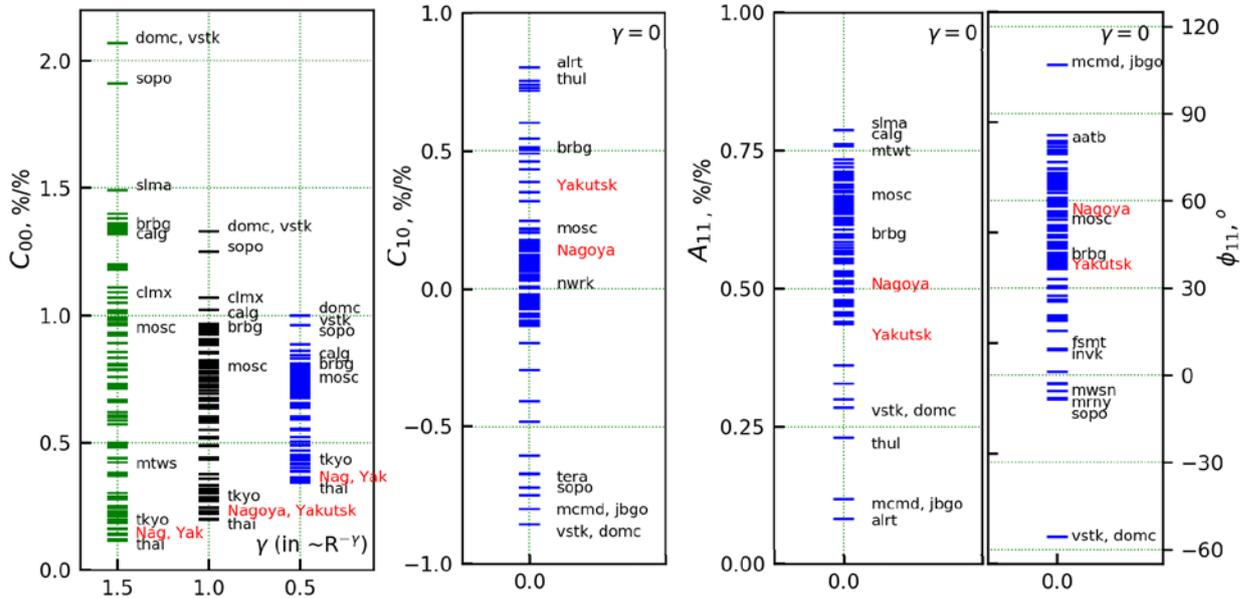

**Figure** 1. Zero harmonic reception coefficients of the CR neutron component for the spectrum indices γ=1.5, 1.0, 0.5 (left panel); north-south CR asymmetry reception coefficients $C_{10}$ (second panel from the left); an amplitude $A_{11}$ of the equatorial component of the anisotropy of the first spherical harmonic and the effective particles drift angle in the Earth's magnetosphere $φ_{11}$ (right panel) according to [Yasue et al., 1982]. For comparison, the reception coefficients of the muon component of two vertical muon telescopes Nagoya and Yakutsk are shown [Fujimoto et al., 1984].

The second panel from the left in **Figure** 1 shows the response of ground-based detectors to the north-south CR asymmetry with its value outside the magnetosphere of 1%. The figure shows that the asymptotic directions of the NM are such that only a few detectors in the Northern and Southern Hemispheres observe more than 50% of the asymmetry. Of these stations, two pairs, Thule-McMurdo and South Pole-Terre Adelie, observe 80% of the asymmetry. The best pair for studying the north-south asymmetry is Thule-McMurdo, since their responses to isotropic variations are almost identical (since they are located near sea level) and they have the weakest response to diurnal variations (the third panel of **Figure** 1), i.e., they are minimally contaminated by isotropic and diurnal variations.

The third panel of **Figure** 1 shows that the equatorial component of the anisotropy of the first spherical harmonic (stellar-diurnal variation) is well measured by the existing network of neutron monitors, since there are ~30 neutron monitors with reception coefficients $C_{11}$>0.5. Only half a dozen polar detectors have a weak response to diurnal variations $C_{11}$<0.3, that is important, as noted above, for identifying north-south asymmetry. To estimate diurnal variation, a uniform distribution of detectors across asymptotic longitude is desirable, which is not accomplished well, especially in the Atlantic and Pacific regions.

## 5. Analisys methods

To solve the approximation problem, at first it is necessary to construct approximation models. The Granitsky–Dorman approximation function $N = c \cdot [1 - \exp(-\alpha R^{-\kappa+1})]$, which proposed in the work [Dorman et al., 1970] has proven to be very effective in processing cosmic ray latitude measurements. It can be expected that the same mathematical model is optimal for approximating the zero harmonic reception coefficients $C_{00}$.

The rigidity dependencies of the first harmonic reception coefficients are more complex, and the approximation models were constructed based on the Granitsky–Dorman function individually. All approximating functions, except for one normalizing parameter $c$, are non-linear. The nonlinear approximation, as well as a comprehensive analysis of the approximation results, was performed using a software package written in Python (https://lmfit.github.io/lmfit-py/model.html#lmfit.model.ModelResult, model "Model().fit()").

## 6. Discussion of results

The approximation results are shown in **Figure**s 2–5 and summarized in Table 1.

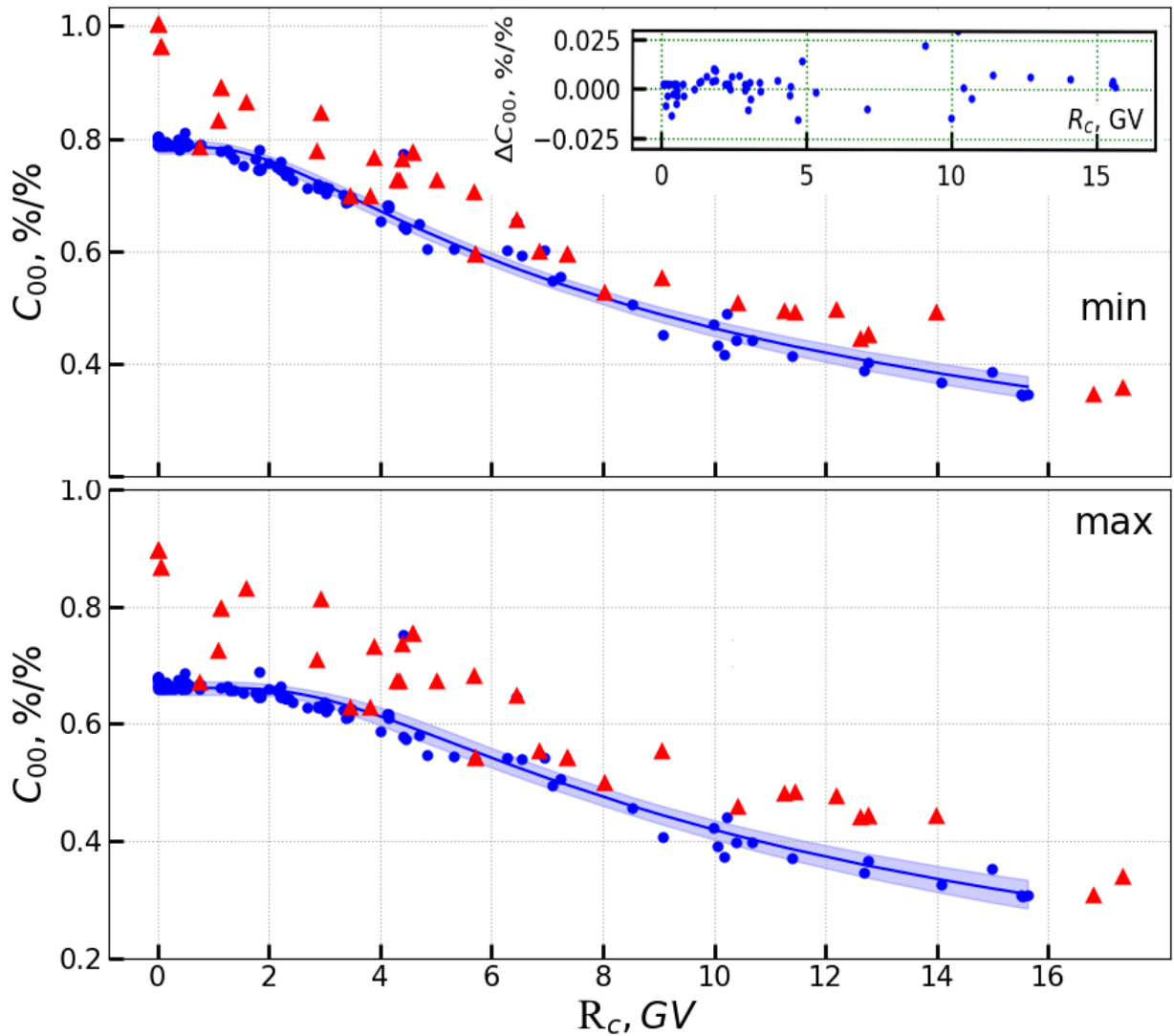

**Figure** 2. Dependence of the zero harmonic reception coefficient $C_{00}$ at the sea level on geomagnetic cutoff rigidity and the confidence interval at the 3σ level. Mountain stations are highlighted with triangles. Reception coefficients $C_{00}$ calculated for the power spectrum index of the zero harmonic γ=0.5 for solar activity minimum and maximum.

**Figure** 2 shows the dependences of the zero harmonic reception coefficients calculated in [Yasue et al., 1982] on the geomagnetic cutoff rigidity for sea level for SA minimum and maximum, and an approximating curve based on the Granitsky–Dorman function.

$C_{00} = c \cdot [1 - \exp(-\alpha R^{-\kappa+1})]$ with 3 parameters estimation. Reception coefficients $C_{00}$ are given for the power spectrum index of the zero harmonic $\gamma=0.5$ and the values of the upper rigidity of CR modulation in the heliosphere $R_U=100$ GV.

In regression analysis, the indicator of the model quality is the coefficient of determination $R^2$, a high value of which indicates an adequate approximation model. In case of $C_{00}$, as it shown in Table 1, $R^2 \approx 0.97$, i.e. 97% is explained by the factors taken into account, and only 3% - by the unaccounted factors. Unexplained part of the coefficient of determination $1-R^2$ can be justified by several reasons. Firstly, the range of depths during the approximation of station data, since mountain-plain stations were used (depth ~ 1000÷900 *hPa*); high-altitude stations, shown in the figure as triangles, were excluded during the approximation. Secondly, the contribution to the unexplained part of the determination coefficient is made by the errors in calculating the coefficients themselves, which are not presented in the work [Yasue et al., 1982]. **Figure** 2 shows the confidence interval at the level 3σ, and for ~100 detectors, only one point may be outside this confidence interval. In reality, up to half a dozen are outside the confidence interval, which is due to the depth range 1000÷900 *hPa* for considered detectors.

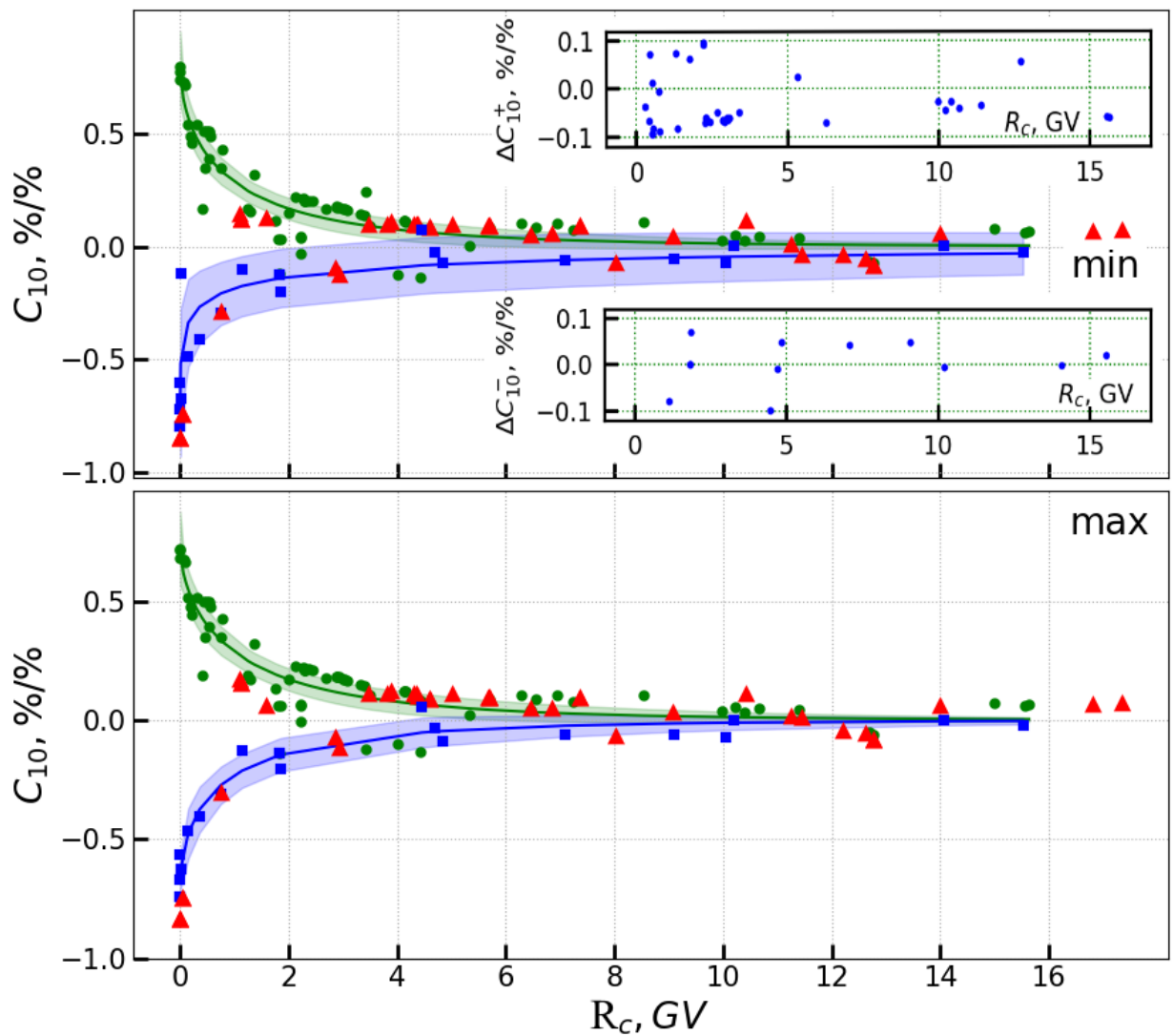

**Figure** *3*. Dependence of reception coefficient $C_{10}$ north-south component of the anisotropy of the first spherical harmonic on the geomagnetic cutoff rigidity and the confidence interval at the level 3σ. Mountain stations is shown as a triangles. Reception coefficients $C_{10}$ calculated for the exponent of the power-law flat spectrum of the first harmonic $\gamma=0.0$, $R_U=100$ *GV* for SA minimum and maximum.

In **Figure** 2, the tab also shows the dependence of the solution's residual on rigidity for SA minimum. Estimation error based on the approximating expression for the reception coefficients $C_{00}$ is about 2%.

Experimental values $\chi_v^2$ are shown in table 1. It is evident that for all the dependency variants under consideration the condition is satisfied $\chi_v^2 < \bar{\chi}_v^2(p)$. Indeed, the tabular value of the critical value of the distribution $\bar{\chi}_v^2(p)$ for the number of degrees of freedom v~90 (number of stations) and for level of trust p=0.01 (significance level 0.99) is $\bar{\chi}_v^2(p)$=137 [https://www.itl.nist.gov/div898/handbook/eda/section3/eda3674.htm].

The special behavior of the north-south asymmetry $C_{10}$ reception coefficients manifests itself in the splitting of the curves into northern and southern branches, determined by the asymptotic directions of the particles. A function based on the Granitsky–Dorman equation was also used as an approximating curve $C_{10} = \pm c \exp(-\alpha R^\kappa)$ for the Northern and Southern Hemispheres with an estimate of three approximation parameters. The results are presented in Table 1. The initial parameters were the reception coefficients $C_{10}$ for the power spectrum index of the zero harmonic γ=0. The graphical result is presented in **Figure** 3, the tab of which also shows the residuals of the performed approximation of the reception coefficients $C_{10}$ for SA minimum. The coefficient of determination $R^2$ for $C_{10}$, as it seems from table 1, is only ≈0.83. This is due to the fact that only a small number of high-latitude stations are actually involved in the analysis; the remaining detectors are not sensitive to these types of variations. The estimation error based on the approximating expression for the reception coefficients $C_{10}$ is about 8 %.

The most difficult task turned out to be the approximation of the reception coefficients of the equatorial component anisotropy amplitude $A_{11}$. The Granitsky–Dorman function with three-parameters estimation was used as an approximating curve $A_{11} = cR^\kappa \exp(-\alpha R^{1/3})$. Multiplier $R^\kappa$ in the form of a power function well reflects the behavior of the function for rigidities in the plateau region of the CR latitudinal plateau region to ~3 *GV*; the exponential decay for rigidities up to the equator poorly reflects the actual behavior of the reception coefficients. This is evidenced by the relatively low coefficient of determination $R^2$=0.71 obtained. It can be seen from **Figure** 4 that exponential decay ends at ~10 *GV*, then the curve tends to a plateau. The results are shown in Table 1. The initial values were the reception coefficients $A_{11}$ for the power spectrum index of the zero harmonic γ=0. Estimation error based on the approximating expression of the reception coefficients $A_{11}$ is about 10%.

Are low values of $R^2$ is always a problem? Regression models with fairly low values $R^2$ (for example, 0.7 for reception coefficient $A_{11}$) may be quite good for the following reason: the found parameter values are average values, but $R^2$ reflect variability, and this ultimately means a large scatter of independent data, and for control, it is necessary to analyze the residual of the mathematical model. As a result, the low coefficient of determination $R^2$ does not diminish the significance of any variables. As a rule, there are no additional reasons to disregard the results found by the model [Frost, 2024].

And, conversely, are high $R^2$ values always reliable? The test is the residual, which should be a random variable with a normal distribution. If any trends appear, this indicates that the linear model is underspecified; the regression line consistently underestimates and overestimates the data along the curve, which is a bias. Such a model must be supplemented with independent variables. Residual plots help ensure the validity of the regression model. [Frost, 2025; Pershin, 2025].

For the dependence of the effective drift angle of particles in the Earth's magnetosphere $\varphi_{11}$ on the geomagnetic cutoff rigidity, one can also use an approximating expression based on the Granitsky–Dorman function $\varphi_{11} = cR^\kappa \exp(-\alpha R)$. The results are shown in Table 1. The initial drift angles of particles in the Earth's magnetosphere were $\varphi_{11}$ for the power-law spectrum index of the zero harmonic $\gamma=0$.

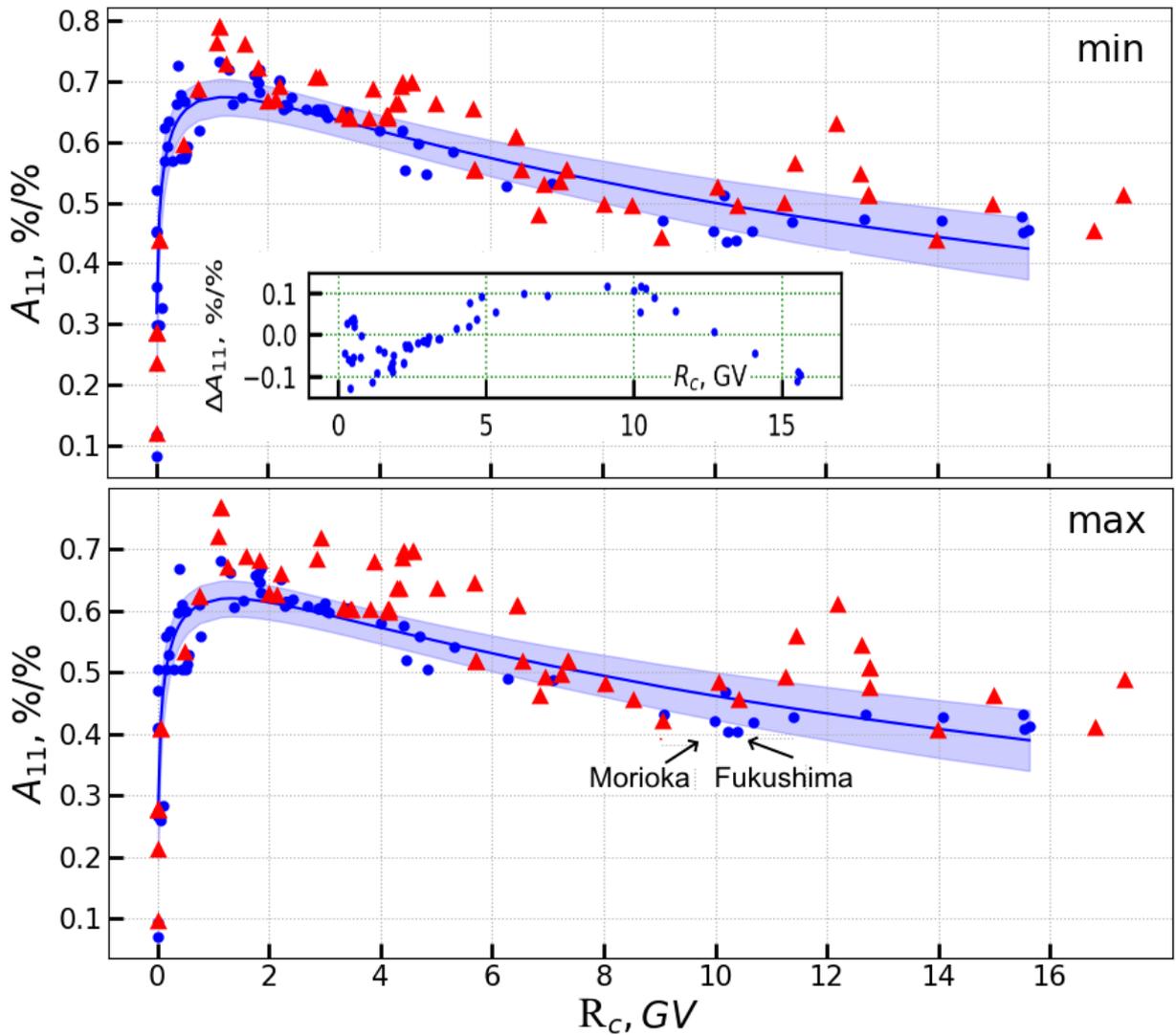

**Figure 4.** Dependence of amplitude of anisotropy equatorial component of the first spherical harmonic $A_{11}$ on the geomagnetic cutoff rigidity and the confidence interval at the 3σ level. Mountain stations are shown as a triangles. Reception coefficients $A_{11}$ calculated for the power spectrum index of the first harmonic $\gamma=0.0$, $R_U=100$ *GV* for SA minimum and maximum.

The result is shown in **Figure 5**, the tab of which also shows the drift angle of particles residuals $\varphi_{11}$ at minimum SA. The coefficient of determination $R^2$ in case of $\varphi_{11}$ (see Table 1) is only ≈0.79. This is due to the fact that only a small number of high-latitude stations are actually involved in the analysis; the remaining detectors are not sensitive to these types of variations. The error in the estimate based on the drift angle approximation $\varphi_{11}$ of particles at SA minimum is about 5%.

Below, in the summary **Figure 6**, estimates of the time (rigidity) dependence of the CR reception coefficients of zero and first harmonics for latitude measurements using the example of the 34[th] expedition of the research vessel Akademik Kurchatov are presented (from December 20, 1981 to April 21, 1982, Vladivostok – Kaliningrad). The vessel's route is shown in the inset of **Figure 6** (upper panel). The expedition was conducted during the SA maximum. During this 4-month period, approximately 50 Forbush decreases of varying amplitude were observed.

During this active SA period, variations of extraterrestrial origin require special attention. Details can be found on the resource https://crst.izmiran.ru/w/feid.

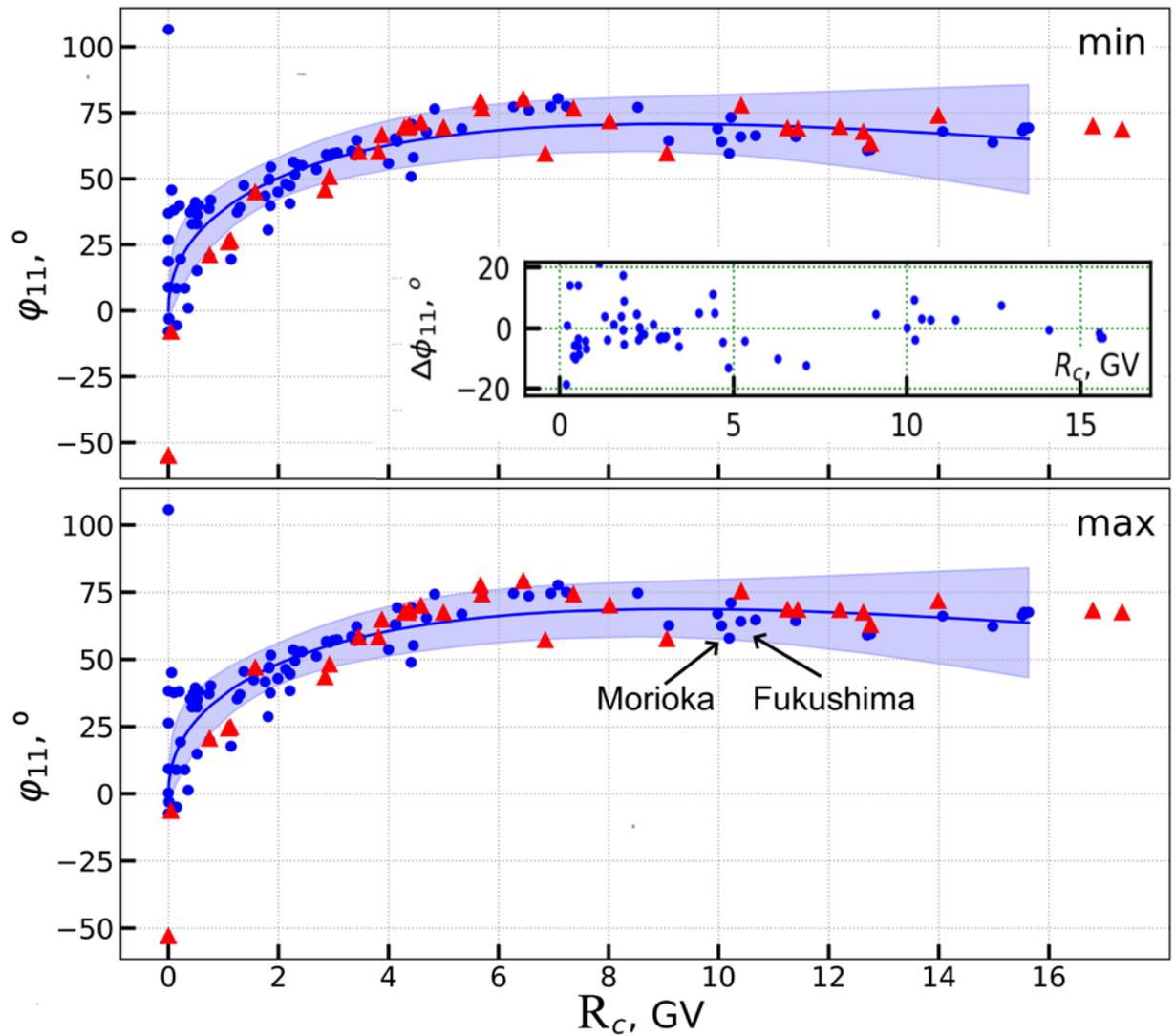

**Figure** *5*. Dependence of the effective drift angle of particles in the Earth's magnetosphere $\varphi_{11}$ on the geomagnetic cutoff rigidity and the confidence interval at the 3σ level. The triangles indicate mountain stations. Reception coefficients $A_{11}$ and effective angle of particle drift in the Earth's magnetosphere $\varphi_{11}$ are calculated for the power spectrum index of the first harmonic γ=0.0, $R_U$=100 *GV* for SA minimum and maximum.

The top panel of **Figure** 6 shows the time dependence of the vertical rigidity of the geomagnetic cutoff rigidity $R_c$ (https://crst.izmiran.ru/cutoff) and the time dependence of the zero harmonic reception coefficients $C_{00}$ along the vessel's route. The reception coefficients were determined based on the results summarized in Table 1. The circles show the reception coefficients of stations that were located nearby. (<2° by latitude and <3° by longitude) of the vessel's route. Due to the unique nature of the route, only a few stations were used: in Japan, Hawaii, the LARC in Antarctica, South America, and several European stations. Agreement is within the error margin.

The second top panel of **Figure** 6 shows the time dependence of the reception coefficient of the CR north-south asymmetry $C_{10}$. It is small, since the ship's route lay in the middle and equatorial latitudes. The third panel of **Figure** 6 shows the time dependance of the equatorial component of the anisotropy amplitude $A_{11}$ and the time dependence of the effective drift angle of particles along the vessel's route. Calculated amplitude values $A_{11}$ and phase $\varphi_{11}$ are within the error range, with the exception of the Japanese stations Morioka and Fukushima. The deviation may be due to a calculation error, which is confirmed for these stations by **Figures** 4 and 5.

**Table 1**. Type of approximation functions and results of approximation of the CR first harmonic reception coefficients for the min. and max. SA.

| Approximation | | min SA | $R^2; \chi^2$ | max SA | $R^2$ |
|---|---|---|---|---|---|
| $C_{00} =$ $c[1-\exp(-\alpha R^{-\kappa})]$ | c | 0.785±0.004 | 0.975 | 0.661±0.004 | 0.948 |
| | α | 6.230±0.360 | 0.001 | 11.15 ±1.29 | |
| | κ | 0.845±0.026 | | 1.045±0.050 | |
| positive branch $C_{10} = +c\exp(-\alpha R^\kappa)$ | c | 0.791±0.052 | 0.828 | 0.722±0.050 | 0.816 |
| | α | 1.021±0.097 | 0.06 | 0.924±0.103 | |
| | κ | 0.603±0.073 | | 0.639±0.084 | |
| negative branch $C_{10} = -c\exp(-\alpha R^\kappa)$ | c | 0.713±0.062 | 0.829 | 0.677±0.025 | 0.965 |
| | α | 1.350±0.235 | 0.03 | 1.075±0.103 | |
| | κ | 0.310±0.090 | | 0.583±0.087 | |
| $A_{11} = cR^\kappa \exp(-\alpha R^{1/3})$ | c | 1.649±0.147 | 0.711 | 1596±0.154 | 0.701 |
| | α | 0.896±0.080 | 0.003 | 0.949±0.087 | |
| | κ | 0.314±0.030 | | 0.343±0.034 | |
| $\varphi_{11} = cR^\kappa \exp(-\alpha R)$ | c | 40.29±1.58 | 0.791 | 38.62±1.52 | 0.803 |
| | α | 0.052±0.010 | 90 | 0.051±0.010 | |
| | κ | 0.468±0.051 | | 0.471±0.051 | |

It should be noted that the Granitskii–Dorman function also describes well the rigidity dependence of the barometric coefficient of the neutron component, which is necessary for processing cosmic ray latitude measurements. The barometric coefficient is determined experimentally for a network of CR stations or directly during expedition work. Experience shows that it is impossible to obtain a reliable series of barometric coefficient values within a single expedition, since at key observation points either sufficient statistics are often not available, or large changes in atmospheric pressure are absent. Consequently, the experimental series of barometric coefficient values is formed based on data from many expeditions. The dependence of the barometric coefficient on rigidity was approximated as

$$\beta(R_c) = \beta_0[1-\exp(-\alpha R_c^{-\kappa})] \qquad (9)$$

for which an approximation coefficients $\beta_0, \alpha, \kappa$ =(0.751%/hPa, 5.69, 0.411) for neutron monitor NM64 and $\beta_0, \alpha, \kappa$ =(0.766%/hPa, 5.69, 0.411) were found for epithermal neutron detector NM64E.

7. **Main conclusions**

1. For all neutron monitors of the global network, the zero and first harmonic reception coefficients were approximated as a function of geomagnetic cutoff rigidity. The numerical calculations from [Yasue et al., 1982], conducted for discrete points in the cosmic ray network, were used as a basis.

2. The developed mathematical models, based on the Granitsky–Dorman function [Dorman et al., 1970], are optimal for approximating the reception coefficients. The type of parameterized functions and the approximation result for the first harmonic reception coefficients and their errors for the SA minimum and maximum are summarized in Table 1. Determination coefficients and criterion $\chi^2$ demonstrate the adequacy of the approximation models used.

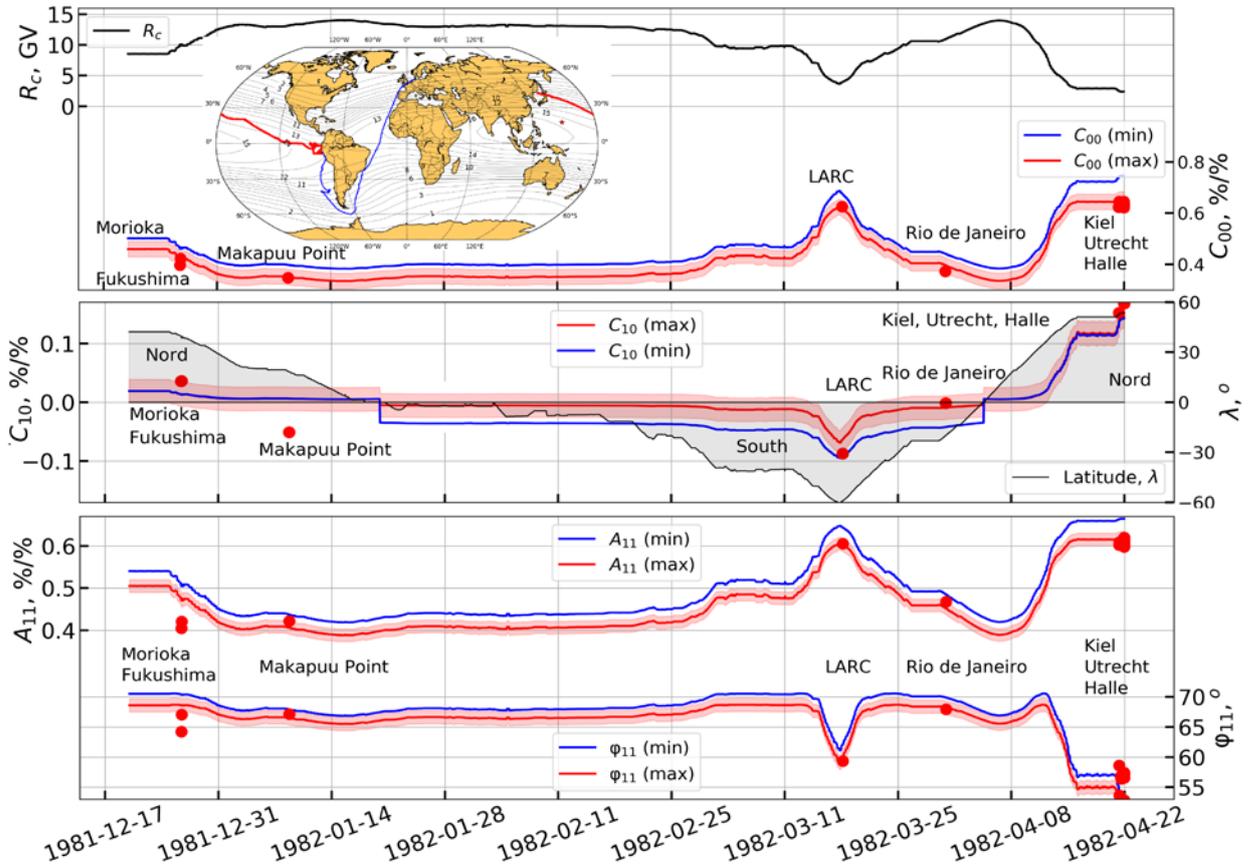

**Figure** 6. Change in the vertical rigidity of the geomagnetic cutoff $R_c$ and the reception coefficients of the zero and first harmonics along the route of the research vessel Akademik Kurchatov from Vladivostok to Kaliningrad. The circles show the reception coefficients of stations that were located nearby (<2° by latitude and <3° by longitude) of the vessel's route.

This is the first time the reception coefficients have been approximated. The results can be applied to the processing of latitude measurements of the cosmic ray neutron component and are also useful for estimating the reception coefficients of the new neutron detectors, of which approximately 20 have been commissioned at this century. In the future, similar work could be carried out for the muon component.

**Acknowledgements**
The authors are grateful to the participants of the NMDB project (www.nmdb.eu). This work is being conducted within the framework of the Russian National ground-based Network of CR Stations (CR Network) (https://ckp-rf.ru/catalog/usu/433536) and the Ministry of Science and Higher Education's (Ministry of Science) Marine Scientific Research project for 2025 and 2027.

**ORCID iDs**
Yanke V.G. (https://orcid.org/0000-0001-7098-9094)
Kobelev P.G. (https://orcid.org/0000-0002-9727-4395)